\documentclass{article}

\begin{document}

\title{The Vast Polar Structure of the Milky Way and Filamentary Accretion of Sub-Halos\footnote{To appear in the proceedings of the Thirteenth Marcel Grossman Meeting on General Relativity, edited by Kjell Rosquist, Robert T Jantzen, Remo Ruffini, World Scientific, Singapore, 2013}}

\author{Marcel S. Pawlowski \\
Argelander Institut f\"ur Astronomie, University of Bonn,\\
Auf dem H\"ugel 71, D-53121 Bonn, Germany\\
E-mail: mpawlow@astro.uni-bonn.de}

\maketitle

\begin{abstract}
The Milky Way (MW) is surrounded by numerous satellite objects: dwarf galaxies, globular clusters and streams of disrupted systems. Together, these form a vast polar structure (VPOS), a thin plane spreading to Galactocentric distances as large as 250 kpc. The orbital directions of satellite galaxies and the preferred alignment of streams with the VPOS demonstrate that the objects orbit within the structure. This strong phase-space correlation is at odds with the expectations from simulations of structure formation based on the cold dark matter cosmology ($\Lambda$CDM). The accretion of sub-halos along filaments has been suggested as the origin of the anisotropic distribution. We have tested this scenario using the results of high-resolution cosmological simulations and found it unable to account for the large degree of correlation of the MW satellite orbits. It is therefore advisable to search for alternative explanations. The formation of tidal dwarf galaxies (TDGs) in the debris expelled from interacting galaxies is a very natural formation scenario of the VPOS. If a number of MW satellites truly are TDGs, mistakenly interpreting them to trace the dark-matter sub-structure of the MW halo would significantly enhance the 'small-scale' problems which are already known to plague the $\Lambda$CDM model.
\end{abstract}



\section{Introduction}\label{sect:intro}
Interpreting the Milky Way (MW) satellite galaxies to be primordial dwarf galaxies which reside in dark matter (DM) sub-halos has unveiled numerous 'small-scale' problems of the currently prevailing $\Lambda$CDM cosmology. These include the cusp/core-problem, the missing satellites problem, the common DM mass-scale of satellite galaxies or the problem of missing bright satellites \cite{Kroupa2012}. Most commonly, these problems are addressed by invoking unaccounted for or unknown baryon physics. The distribution of sub-halos around a host galaxy like the MW, however, is essentially independent of the detailed baryon physics because the gravity exerted by DM dominates on scales of tens to hundreds of kpc. 
If satellite galaxies trace the DM sub-halo population, their positions in phase-space can be used to test cosmological models, which predict a near-isotropic distribution.

\section{The VPOS: a vast polar structure around the Milky Way}\label{sect:vpos}
The observed distribution of different satellite objects of the MW is strongly anisotropic \cite{Lynden-Bell1976,Kroupa2005}. They lie in a vast polar structure (VPOS) \cite{Pawlowski2012a}, a thin planar distribution that is oriented perpendicular to the MW disc. This VPOS consists of:
\begin{itemize}
\item The 11 'classical', bright MW satellite galaxies. The root-mean-square (RMS) height of the satellite galaxies from the best-fitting plane describing their positions is only 18.5 kpc, while they have radial distances of up to 250 kpc.
\item The 13 recently discovered faint satellites. A plane fitted to only these objects is oriented like that of the 'classical' satellite galaxies \cite{Kroupa2010} and has a RMS height of 28.6 kpc. While most faint satellites have been discovered in the SDSS, the incomplete sky coverage is not responsible for the alignment because within the survey region there is a clear deficit of satellites far off the VPOS \cite{Metz2009}.
\item The 'young halo' globular clusters (GCs), which have ages in the range of 9 to 11 Gyr. Their best-fitting plane has a RMS height of 11.8 kpc and is less than $13^{\circ}$\ inclined with respect to the plane fitted to all 24 satellite galaxies.
\item The stellar and gaseous streams in the MW halo, originating from disrupted satellite galaxies and GCs. Seven out of 14 known streams align with the VPOS. As the orientation of the streams trace the orbital planes of their parent objects, this indicates that the MW satellite objects preferentially orbit within the VPOS.
\item The orbits of the MW satellite galaxies. The orientation of the angular momentum directions can be determined \cite{Metz2008} for now 11 MW satellite galaxies for which proper motions have been measured. These indicate that, within the uncertainties, nine out of the 11 satellites orbit within the VPOS. All but one of these nine co-orbit in the same direction. The VPOS is rotating and is thus not only a spatial, but in fact a phase-space structure.
\end{itemize}

\section{Filamentary Accretion of Sub-Halos as the Origin of the VPOS}\label{sect:filament}
The accretion of DM sub-halos along cosmic filaments has been suggested to give rise to VPOS-like, planar distributions of coherently rotating satellite galaxies \cite{Libeskind2011,Lovell2011}.
However, the typical size of DM filaments, which have radii on the order of 500 to 1000 kpc, is much larger than the virial radius of the presumed DM halo of the MW ($\approx 250$~kpc), and more than two orders of magnitude larger than the typical distance of the MW satellites from the best-fitting VPOS plane ($< 30$~kpc). This size difference results in a near-isotropic accretion onto the main halos \cite{Vera-Ciro2011}, excluding the filamentary accretion scenario as an explanation of the VPOS.

In addition to this purely spatial argument, the velocity information can be used in a more thorough investigation of the suggested filamentary accretion scenario. We have developed a test which compares the distribution of angular momentum directions of the MW satellites with that of modelled data. This test was applied to the DM sub-halos formed in eight high-resolution cosmological simulations of MW-like halos \cite{Pawlowski2012b}. Our main findings are that the cosmological simulations are extremely unlikely to reproduce the observed clustering of angular momentum directions and their closeness to a polar orbit. For most simulated halos the likelihood is less than 0.5\%, only the Via Lactea II simulation gives a higher likelihood of 1.5\%. These values are comparable to drawing from an isotropic distribution (0.5\%). 

\section{Implications for the Interpretation of the MW Satellite Galaxies}\label{sect:concl}
When using the full phase-space information of the MW satellites (i.e. the anisotropy in positions \textit{and} velocities), and adding the information contained in the other VPOS-related objects (YH GCs, streams), it becomes apparent that a purely cosmological accretion can not explain the significant structuring of the MW satellite distribution. 
We therefore suggest a different origin for the VPOS: the formation of tidal dwarf galaxies (TDGs) from the debris of interacting galaxies \cite{Pawlowski2011, Pawlowski2012a}. TDGs (and also GCs) form in sufficient number, can be long-lived and are correlated in phase-space (see the detailed discussion in Ref. \cite{Pawlowski2012a}). Applying our test to modelled data of tidal debris demonstrates that these can be up to a factor of 100 more likely to reproduce the observed satellite galaxy angular momentum distribution than the cosmological state-of-the-art simulations \cite{Pawlowski2012b}. 

In the TDG scenario, the MW had an encounter with another galaxy about 10 Gyr ago, leading either to a merger or a fly-by \cite{Pawlowski2011, Pawlowski2012a}. Alternatively, it has been suggested that TDGs formed in a merger involving M31, which were then expelled into the direction of the MW where they now arrive to form the VPOS \cite{Fouquet2012}.
Within current observational bounds, the TDG-approach is well consistent with the MW satellite population and is the only currently known model which can naturally explain the strong phase-space correlation of the VPOS. When interpreting the satellite galaxy population of the MW, one therefore needs to be aware that a significant fraction of it might consist of ancient TDGs \cite{Dabringhausen2012}. Spuriously interpreting all MW satellites to trace dark-matter sub-halos will then give rise to inconsistencies with the cosmological predictions and contribute to the numerous 'small-scale' problems which plague the current $\Lambda$CDM standard model of cosmology. If this is the case, adjusting the cosmological model in order to resolve such problems could hamper the development of an optimal cosmological description of the Universe. It is therefore of uttermost importance to determine the degree of TDG-'contamination' present in the local dwarf galaxy population.

\bibliography{PawlowskiBib}

\end{document}